\documentclass{article}

\usepackage{moreverb,url}

\usepackage[colorlinks,bookmarksopen,bookmarksnumbered,citecolor=red,urlcolor=red]{hyperref}

\usepackage{tikz}
\usepackage{listings}
\usepackage{siunitx}
\usepackage{makecell}
\usepackage{todonotes}
\usepackage{nameref}
\usepackage{cleveref}
\usepackage[table]{xcolor}

\usepackage{amssymb}

\newcommand\BibTeX{{\rmfamily B\kern-.05em \textsc{i\kern-.025em b}\kern-.08em
T\kern-.1667em\lower.7ex\hbox{E}\kern-.125emX}}

\newcommand{\orcid}[1]{\href{https://orcid.org/#1}{\includegraphics[height=10pt]{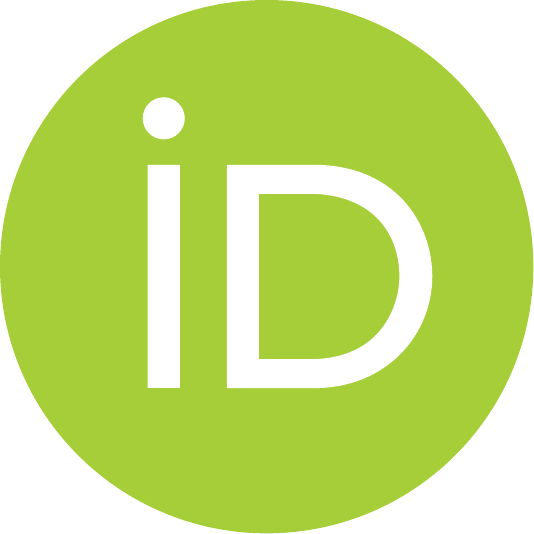}}}

\usepackage{booktabs}

\usepackage{arxiv}
\begin{document}


\title{Asynchronous-Many-Task Systems: Challenges and Opportunities - Scaling an AMR Astrophysics Code on Exascale machines using Kokkos and HPX}

\author{Gregor Dai\ss\orcid{0000-0002-0989-5985}, Alexander Straub\orcid{0000-0002-6749-9710}, Dirk Pfl\"uger\orcid{0000-0002-4360-0212}\\University of Stuttgart, Stuttgart, 70569 Stuttgart, Germany  \AND 
    Patrick Diehl\orcid{0000-0003-3922-8419}, Christoph Junghans\orcid{0000-0003-0925-1458}\\
    Los Alamos National Laboratory, Los Alamos, NM, 87545 U.S.A.\AND
    Jiakun Yan\orcid{0000-0002-6917-5525}\\University of Illinois Urbana-Champaign, Champaign, IL, 61801 U.S.A. \AND 
    John K. Holmen\orcid{0000-0002-5934-2641}\\  Oak Ridge National Laboratory, Oak Ridge, TN, 37831 U.S.A. \AND  Rahulkumar Gayatri\\Lawrence Berkeley National Laboratory, Berkeley, CA 94720 U.S.A.  \AND 
    Jeff R. Hammond\orcid{0000-0003-3181-8190}\\NVIDIA Helsinki Oy, Helsinki, 00180 Finland \AND
    Dominic Marcello\orcid{0000-0002-2782-315X}, Hartmut Kaiser\orcid{0000-0002-8712-2806}\\ Louisiana State University, Baton Rouge, LA, 70803 U.S.A.\AND Miwako Tsuji\orcid{0000-0003-4709-1969}\\ RIKEN Center for Computational Science, Kobe, 650-0047 JAPAN
      }




\maketitle

\begin{abstract}\footnotesize
Dynamic and adaptive mesh refinement is pivotal in high-resolution, multi-physics, multi-model simulations, necessitating precise physics resolution in localized areas across expansive domains. Today's supercomputers' extreme heterogeneity presents a significant challenge for dynamically adaptive codes, highlighting the importance of achieving performance portability at scale. Our research focuses on astrophysical simulations, particularly stellar mergers, to elucidate early universe dynamics. 
We present Octo-Tiger, leveraging Kokkos, HPX, and SIMD for portable performance at scale in complex, massively parallel adaptive multi-physics simulations. Octo-Tiger supports diverse processors, accelerators, and network backends. 
Experiments demonstrate exceptional scalability across several heterogeneous supercomputers including Perlmutter, Frontier, and Fugaku, encompassing major GPU architectures and x86, ARM, and RISC-V CPUs. Parallel efficiency of $47.59$\% ($110,080$ cores and $6880$ hybrid A100 GPUs) on a full-system run on Perlmutter ($26$\% HPCG peak performance) and $51.37$\% (using $32,768$ cores and $2,048$ MI250X) on Frontier are achieved.
\end{abstract}

\keywords{Adaptive mesh refinement, Asynchronous-many-task systems, Kokkos, HPX, Exascale computing, Stellar merger, High performance computing}

\cleardoublepage
\newpage
\twocolumn

\section{Introduction}
\begin{figure}
    \centering
    \newcommand{\colbox}[1]{\protect\tikz{\protect\draw[#1,fill=#1] (0,0) rectangle (0.2,0.2);}}%
    \definecolor{isoblue}{RGB}{1,142,217}%
    \definecolor{isoorange}{RGB}{255,170,0}%
    \definecolor{isogreen}{RGB}{0,199,0}%
    \definecolor{isoyellow}{RGB}{239,239,53}%
    \definecolor{isoviolet}{RGB}{170,85,255}%
    \definecolor{isored}{RGB}{200,0,0}%
    \begin{tikzpicture}%
        \node[anchor=south west,inner sep=0] at (0,0) {%
            \includegraphics[trim={0 0 0 4cm},clip,width=\linewidth]{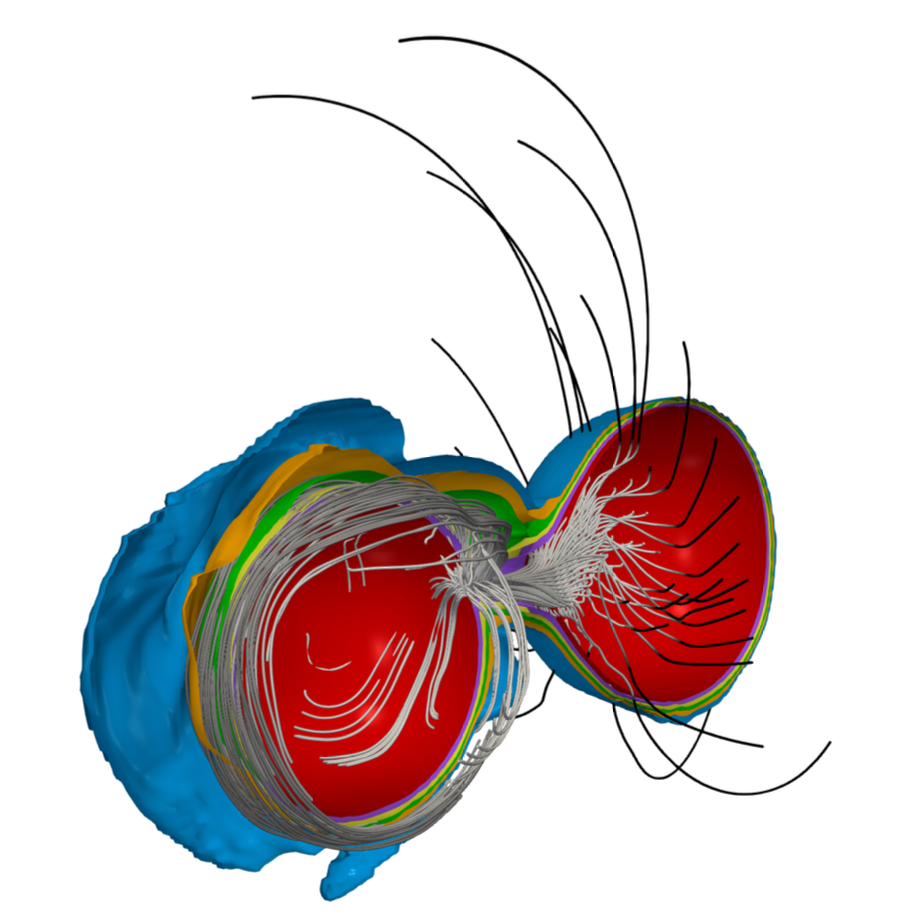}%
        };%
        \node[anchor=north west,inner sep=0] at (0,6.6) {%
            \begin{minipage}{2cm}
                \colbox{isoblue}~$\rho=0.1$ \\
                \colbox{isoorange}~$\rho=1$ \\
                \colbox{isogreen}~$\rho=5$
            \end{minipage}
            \begin{minipage}{2cm}
                \colbox{isoyellow}~$\rho=20$ \\
                \colbox{isoviolet}~$\rho=100$ \\
                \colbox{isored}~$\rho=500$
            \end{minipage}
        };%
        \node[anchor=north west,inner sep=0] at (0,5.2) {%
            in $g/cm^3$
        };%
    \end{tikzpicture}%
    \caption{The figure shows an exploded view of the internal structure of a double white dwarf binary star system during the mass transfer phase that precedes the eventual merger of the two stars into a single object. Color is used to indicate density layers ($0.1$ to $500$ $g/cm^3$). The star to the right of the image (the donor) is transferring mass through a stream striking the upper layers of the more massive white dwarf to the left (the accretor). Several pathlines~(gray) show how the stream impacts the accretor, with most of the mass flowing around the accretor while some fraction of the stream spreads in all directions around the point of impact.}
    \label{fig:dwd_merger}
    \vspace{-1.5em}
\end{figure}

Adaptive mesh refinement (AMR) is a crucial component for many high-resolution, multi-physics, and multi-model simulations. 
In this work, we are focusing on one such simulation code: the astrophysics application Octo-Tiger.
Octo-Tiger simulates binary star systems\textcolor{black}{, allowing one to study} their interactions and the various phenomena caused by them.
In these simulations, large computational domains are needed, for example, to track the mass leaving the stars; see the streamlines in Figure~\ref{fig:dwd_merger}. To resolve the quantity of interest, such as the concentration of ${}^{16}\textrm{O}$, some areas of the large computational domain, like the stars (donor and accretor) and accretion column, require a high resolution, which is why AMR is such a crucial component.
To provide an example of such an adaptively refined mesh with Octo-Tiger,
Figure~\ref{fig:octo-tiger-3d-mesh} shows the mesh close to the merger on the equatorial plane. The color shows the portion of the grid assigned to various computational nodes. The workload of the nodes is unbalanced \textcolor{black}{with respect to} the domain\textcolor{black}{. This results in domains of differing sizes being allocated across} nodes to achieve load balancing across the mesh. In addition, some nodes have more cells due to the refined mesh; where the smallest cells have a size of \num{1.4e7} cm and the largest cells have a dimension of \num{2.24e8} cm. 

This highlights the various scalability challenges we face from the High-Performance-Computing (HPC) perspective when targeting large production simulations on supercomputers with such AMR codes: load-balancing and memory management are fundamental challenges, for example.
Another crucial challenge is dealing with the irregular parallelism: AMR implementations usually involve a tree-based data-structure, which poses difficulties when trying to exploit the available parallelism efficiently. Here, we first have to traverse the tree for the work to become available within the system. Especially in large distributed applications, this can lead to starvation of the used supercomputer.
Given the diverse set of current supercomputers (including machines with accelerators from NVIDIA, AMD, and Intel) portability is also a concern when developing these large simulation codes.

Given how widespread these concerns are for developing scalable AMR applications, there are existing solutions available to address them.
Specifically, for our work, we turn to HPX and Kokkos.
HPX, a distributed asynchronous task-based runtime system, can help us to tackle scalability. Expressing the parallelism within Octo-Tiger's tree-structure with the dynamic HPX task-graph and utilizing HPX's distributed features helps with many of the aforementioned challenges, for instance easing the development effort for the distributed memory management, transparent overlapping of computation and communication, and exploiting the parallelism during Octo-Tiger's solver iterations.
Kokkos, a framework for developing portable compute kernels, in turn helps us with portability in terms of code and performance.
Thus, in theory, combining HPX and Kokkos presents an opportunity to develop distributed and portable HPC applications more easily, helping to achieve scalability on a wide range of CPU and GPU supercomputers.

However, in practice, we have found that this approach required some additional glue in order for the frameworks to work together seamlessly when developing an AMR application, such as Octo-Tiger. Hence, we needed to develop some integrations and add missing pieces and tools where necessary.
In some scenarios, the frameworks themselves did not work \textcolor{black}{well} together (such as Kokkos' internal fences (barriers) needlessly blocking HPX worker threads instead of suspending the HPX task). 
\textcolor{black}{We also} encountered problems regarding missing tools to adapt our use-case to the given hardware. For example, Octo-Tiger was originally developed only for CPUs. When porting Octo-Tiger to Kokkos, the workload per tree-node proved too small to provide sufficient work for an efficient compute kernel (especially on GPUs), causing poor performance due to GPU device starvation. This is a problem shared by many tree-based codes. Moreover, the dynamic nature of the work (i.e., tree-nodes being added, removed, and migrated over the course of the simulation) greatly complicates any static approach for aggregating the workload of different tree-nodes to address this.

Over the recent years, we faced and addressed many of these issues and missing pieces in our previous work. Notably, we aimed to do so in a way that makes our solutions usable universally in codes other than Octo-Tiger.
\textcolor{black}{As a part of past work, we} improved the interoperability of HPX and Kokkos considerably, allowing us to directly and asynchronously integrate Kokkos operations into the HPX task-graph, eliminating the need to block CPU threads with blocking fence operations entirely~\cite{9460406}.
We addressed the device starvation issue by implementing a set of allocators and executors that facilitate dynamic work aggregation, combining individual kernel launches of compatible kernels on-the-fly into a single large compute kernel guided by the current GPU work load~\cite{10024622}.
Furthermore, we made use of SIMD types within our Kokkos kernels (allowing for both the Kokkos SIMD types and the \texttt{std::simd} types), to improve the efficiency on CPU platforms whilst seamlessly retaining GPU support. Here, we contributed \texttt{std::simd}-compatible SIMD types for Scalable Vector Extension (SVE)~\cite{10.1109/P3HPC56579.2022.00014} and RISC-V Vector Extension (RVV)~\cite{diehl2024preparinghpcriscvexamining} to improve platform support, especially for Fugaku~\cite{diehl2024fugaku}.
We also added an improved networking backend to HPX~\cite{jiakun2023hpxlci}.
Each of these previous works considerably improved the usability of HPX and Kokkos for developing scalable and portable AMR applications, thus helping us to realize the opportunities that the combined usage of HPX and Kokkos offers.
Over the course of this previous work, we also ported Octo-Tiger's most important compute kernels to Kokkos, turning it into a GPU-accelerated application.

In this work, we build on these achievements, combine all methodologies, enable the seamless exchange of different backends on cutting-edge systems, and show novel scalability results up to full system runs for three of the world's fastest supercomputers with entirely different hardware.
Thus, the work presented here builds directly on our previous work, combining those results in Octo-Tiger to \textcolor{black}{allow one to run} it on multiple major supercomputers and examine the scalability and performance of production scenarios at scale on these systems.

Specifically, the contributions of this work include runs on Perlmutter (up to the full system), Frontier (up to 1024 compute nodes), and Fugaku (up to 6144 compute nodes). On Perlmutter, we also include tests using different HPX networking backends. In order to underline the achieved wide portability of Octo-Tiger, we further provide relevant runtime data on several other compute architectures.
To us, these runs and the generated runtime data provide useful information regarding where we stand performance-wise with Octo-Tiger and what our next steps should be. To the reader, this provides a real-world use-case of a high-resolution, multi-physics AMR code being scaled to multiple machines using HPX and Kokkos.

Additionally, in this work, we provide a better overview of all of our previous results to make HPX and Kokkos more suitable for applications such as Octo-Tiger.
This is to provide the reader with a better insight into what features and framework integrations we had to implement to scale our AMR application to GPU supercomputers in the first place. Crucially, these features and integrations all work independent of Octo-Tiger and are available for other HPC application developers in case they consider using HPX and Kokkos.
\textcolor{black}{As a result,} this overview may also prove to be useful for developers of frameworks similar to HPX: it outlines what challenges need to be overcome for efficient use of a distributed task-based framework with Kokkos, especially when dealing with GPUs and tree-based user codes.

The remainder of this work is structured as follows:
First, we will mention related work in the next section.
Then, as Octo-Tiger is a major part of this work, serving both as motivation and as a benchmark for many of our developments, we will cover its astrophysical use-case and the computational challenges in greater detail afterwards.
Next, we will introduce HPX, Kokkos, and Octo-Tiger themselves.
Then, we will provide the overview of our previous work that addressed the challenges we found when using HPX and Kokkos in a distributed AMR application. 
After that, we move to describing the distributed runs we performed on various machines with Octo-Tiger, realizing the opportunities presented by combining HPX and Kokkos, namely scalability and portability.
Lastly, given the runtime data collected in the previous section, we discuss our next steps to improve Octo-Tiger and conclude this work.
 
\section{Related Work}
\label{sec:related:work}

Concerning the astrophysics application, the CASTRO~\cite{almgren2020castro} application supports adaptive mesh refinement using AMReX~\cite{doi:10.1177/10943420211022811} with similar astrophysics features. Castro uses the \textit{MPI+X} approach where \textit{X} is OpenMP for CPUs and \textit{X} is either CUDA for NVIDIA GPUs or \textcolor{black}{HIP} for AMD GPUs. Octo-Tiger and CASTRO are both astrophysical simulation codes, but they differ in numerical approach and intended use cases. CASTRO is a general-purpose, block-structured AMR code designed for a broad range of compressible flow problems, employing finite volume hydrodynamics and multigrid gravity solvers. Octo-Tiger, in contrast, is specialized for modeling rotating binary star systems. It uses an octree-based AMR mesh with finite volume hydrodynamics that uses the piecewise parabolic method (PPM) to compute the values at all 26 spatial directions on a grid cell. It includes a Self-Consistent Field (SCF) solver capable of generating synchronously rotating binary systems in equilibrium. This is a class of initial conditions that CASTRO cannot natively produce. Octo-Tiger also employs a Fast Multipole Method (FMM) for gravity, including the time derivative of the potential to ensure machine-precision energy conservation in the rotating frame. These differences in both software and scientific focus mean that direct performance comparisons between the two codes are not meaningful, as they are solving fundamentally different problems. Table~\ref{tab:physics:features} showcases the supported physics modules. Castro also provides nuclear reaction networks and thermal diffusion, which are not yet available in Octo-Tiger. Other MPI-based solutions are Paramesh~\cite{macneice2000paramesh}, SAMRAI~\cite{wissink2001large}, FLASH~\cite{fryxell2000flash}, p4est~\cite{burstedde2011p4est}, Enzo~\cite{o2005introducing},
Chombo~\cite{colella2012chombo}, and deal.II~\cite{bangerth2007deal}. Focusing on oct-tree based adaptive mesh refinement, tree traversals are needed and depending on the level of refinement $d$ this takes $\mathcal{O}(d)$ global reductions. For large scale simulations and high level of refinements, synchronization and communication overheads can affect the efficiency~\cite{langerhighly}. Potential solutions are \textit{1)} to use non-blocking MPI calls, \textit{2)} overlap communication and communication, and \textit{3)} load balancing. The asynchronous launches of HPX tasks easily facilitate non-blocking calls, describing tasks dependencies using HPX's task graph allows for a natural overlap of communication and computation, and HPX's support for work stealing of tasks between threads within a node helps with load balancing. A similar approach has been done using Charm\texttt{++}~\cite{langerhighly}.

\begin{table}[tb]
    \centering
    \rowcolors{2}{gray!25}{white}
    \begin{tabular}{l|cc}\toprule
    Method & Castro & Octo-Tiger  \\\midrule
    Radiation     & \checkmark & \checkmark \\
    Gravity & \checkmark & \checkmark \\
    Hydro & \checkmark & \checkmark \\
    Nuclear reaction networks & \checkmark & -- \\
    Thermal diffusion & \checkmark & -- \\\bottomrule
    \end{tabular}
    \caption{Comparison of the physics modules available in Octo-Tiger and Castro. }
    \label{tab:physics:features}
\end{table}

From the code portability perspective, the following alternatives to Kokkos are available: \textcolor{black}{OCCA}, OpenACC, OpenCL, RAJA~\cite{beckingsale2019raja}, and SYCL. 
We decided to use Kokkos for the two following reasons: \textit{1}) we wanted to use a C\texttt{++} library since HPX leverages the C\texttt{++} standard and that would simplify the development of our integrations and \textit{2}) AMD, NVIDIA, and Intel GPU support were strong requirements. 
For the GPUs, we use the respective Kokkos GPU execution spaces. HPX is not the only asynchronous many-task system (AMT) with GPU capabilities in existence. Since this work focuses on distributed runs, we will mention only distributed AMTs. Table~\ref{tab:amt:gpu} shows the GPU support of Chapel~\cite{chamberlain2011chapel}, Charm\texttt{++}~\cite{kale1993charmpp}, Legion~\cite{bauer2012legion}, Uintah~\cite{germain2000uintah}, and PaRSEC~\cite{bosilca2013parsec}. For a detailed survey, we refer to~\cite{thoman2018taxonomy}. Only Legion and Uintah have Kokkos support. Uintah adopts Kokkos through a Uintah-specific intermediate portability layer used to preserve legacy code and centralize Kokkos calls~\cite{sunderland2016overview}. Legion allows for Kokkos API calls. To conclude, the novelty of our work is that Legion and Uintah solely use \textcolor{black}{Kokkos'} GPU execution space while we additionally use Kokkos' HPX backend to launch compute kernels on the CPU.

\begin{table*}[tb]
    \centering
    \rowcolors{2}{gray!25}{white}
    \begin{tabular}{c|ccc|ccc}\toprule
     & \multicolumn{3}{c|}{GPU support} & \multicolumn{3}{c}{Kokkos}  \\\midrule
    AMT     & NVIDIA & AMD & Intel & 
    \texttt{Kokkos::OpenMP}  &  \texttt{Kokkos::HPX} & \texttt{Kokkos::CUDA/HIP/SYCL} \\\midrule
    Chapel     & \checkmark & \checkmark & & & &\\ 
    Charm\texttt{++}     & \checkmark & & \\ 
    Legion     & \checkmark & \checkmark & \checkmark & \checkmark & & \checkmark\\ 
    Uintah &  \checkmark & \checkmark & \checkmark & \checkmark & &  \checkmark\\
    PaRSEC &  \checkmark & & & & & \\
    HPX &  \checkmark & \checkmark & \checkmark &  & \checkmark & \checkmark \\\bottomrule
    \end{tabular}
      \caption{Overview of GPU architecture support for various asynchronous many-task systems (AMT). Only HPX, Legion, and Uintah support Kokkos. All of them use \texttt{Kokkos::Cuda}, \texttt{Kokkos::HIP}, \texttt{Kokkos:SYCL} to launch the respective GPU kernels. For the kernel launches on the CPU, Legion and Uintah use the \texttt{Kokkos::Serial} and \texttt{Kokkos::OpenMP} execution spaces. Only HPX provides a Kokkos CPU backend \texttt{Kokkos::HPX} to run Kokkos kernels on HPX threads.}
    \label{tab:amt:gpu}
\end{table*}

\section{Overview of the problem}
\label{sec:problem}
\subsection{The Astrophysical Problem}
\label{sec:astro_problem}
The majority of stars in the Universe are not isolated stars but, rather, members of gravitationally bound systems of stars with two or more components. Binary stars are star systems with two components that form 
from the same cloud of gas. 
The components of binary systems may eventually evolve to the ``white dwarf" stage, where nuclear burning has ceased, much of the stellar material has been ejected, and remaining remnants emit only black-body radiation. Known as a double white dwarf (DWD), the components may be driven closer together through emission of gravitational radiation.
When the components of a DWD are close enough that the gravity of the ``accretor" pulls mass from the ``donor," this is said to be an ``interacting" DWD (see Figure~\ref{fig:dwd_merger}). Interacting binary systems are commonplace in the Universe. A large fraction of interacting binaries result in the disruption of the donor and subsequent merger of much of its material with the accretor, which makes understanding them crucial to our understanding of a wide variety of potential outcomes for such mergers (\emph{e.g.}, type 1A super-novae, R Coronae Borealis stars).

The R Coronae Borealis stars have strange elemental abundances \textcolor{black}{as they are} low in hydrogen but high in carbon content. They are variable stars, fading from and then returning to maximum brightness at intervals of several years. This is thought to be caused by clouds of carbon dust surrounding the star. They are thought to 
form from the merger of double white dwarfs. Octo-Tiger has been used extensively in the past to study this question. \emph{e.g.}\ ~\cite{10.1093/mnras/stab937,shiber2024hydrodynamic,Staff_2012,Lauer2019}. Octo-Tiger has also been used to investigate whether the star Betelgeuse may be the outcome of a past merger ~\cite{Manos2020}.

The numerical modeling of the short-lived phase of unstable mass transfer in a DWD merger requires a fully three-dimensional, self-consistent treatment of the effects of hydrodynamics and gravity in a rotating frame of reference. There are two basic approaches to the hydrodynamics: Smoothed Particle Hydrodynamics (SPH) and the grid-based finite volume method (FVM). One of the processes crucial to understanding the formation of R Coronae Borealis stars is the creation of an unusually high concentration of ${}^{16}\textrm{O}$
in the accretor during merger. SPH codes tend to greatly underestimate this effect, making it difficult to resolve the low mass transfer rates typical of most of the merger processes. FVM codes tend to overestimate it. With FVM, however, it is possible to conduct convergence studies by successively doubling grid resolution to better understand the magnitude of the overestimation.

\begin{figure}[tb]
    \centering
    \includegraphics[trim={2cm 1cm 0 1cm},clip,width=\linewidth]{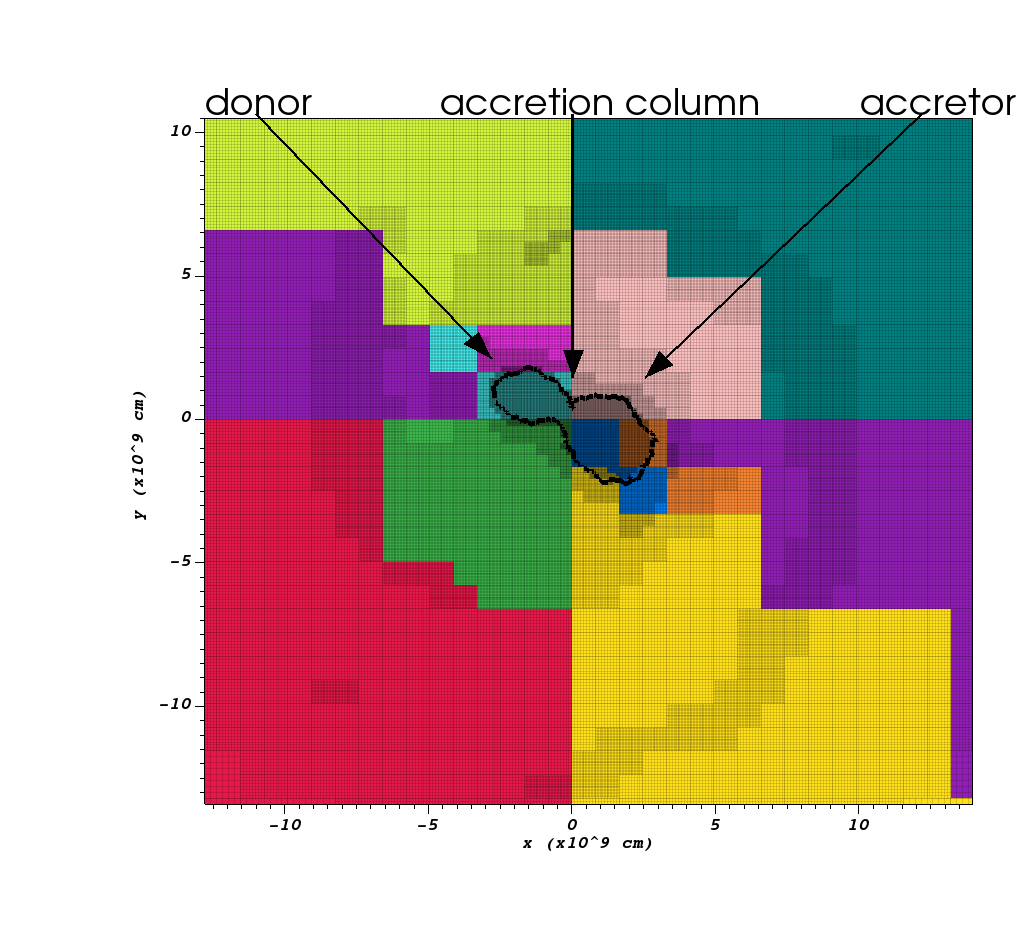}\\[-2em]
    \caption{The adaptively refined grid close to the merger for level 11 on an equatorial plane. The contour of the donor star and accretor star are sketched in black. Between the two stars is the accretion column where the mass transfer between the stars occurs. The color shows the portion of the grid assigned to different nodes. The figure is zoomed in on the two stars and does not show the coarser grid further away from the stars.}
    \label{fig:octo-tiger-3d-mesh}
    \vspace{-1.5em}
\end{figure}

\subsection{The computational challenges}
\label{sec:comp_problem}

The need for application codes to be portable---both in terms of code and performance---is one of the central problems of modern HPC computing. 
The quickly growing complexity of hardware architectures and software stacks require rapid advancements in development productivity, performance, and software portability. This work relies on standard C\texttt{++} and widely used portable C\texttt{++} libraries, such as Kokkos (a C\texttt{++} Performance Portability Programming Ecosystem~\cite{9485033,CarterEdwards20143202}), HPX (a fully C\texttt{++} standards conforming asynchronous many-task runtime system~\cite{10.1145/2676870.2676883,kaiser2020hpx,kaiser_hartmut_2023_8216176}), and C\texttt{++} standard SIMD features that have been added to C\texttt{++}26~\cite{par_ts_standard} to help ensure full code portability of our application. Here, we used an early version of the SIMD features (\texttt{std::experimental::simd}) to allow us to stay compatible with older C\texttt{++} versions (Octo-Tiger can still be compiled using C++20 compilers).
%
The use of Kokkos, HPX, and SIMD Features guarantees its seamless portability and consistent scaling characteristics across a wide variety of heterogeneous (GPU and non-GPU based) hardware platforms, in particular on Intel, AMD, and NVIDIA products, but also on upcoming ARM64 and RISC-V HPC architectures.

In the problem domain investigated, the amount of computational work per compute kernel is low, requiring adaptive work aggregation on GPU devices to reduce scheduling overheads and starvation effects for the related asynchronous tasks~\cite{10024622}. The nature of highly complex 3D AMR applications like Octo-Tiger (see Figure~\ref{fig:octo-tiger-3d-mesh}) require exascale computing systems to solve underlying physics problems that require very large amounts of data and pose computational challenges. To achieve high parallel efficiency, these challenges include providing intra- and inter-node load-balancing, pervasive overlapping of computation and communication, tight integration and high utilization of accelerator devices, and runtime-adaptive task placement and scheduling across nodes and diverse execution environments.

\section{Frameworks and Application}
\label{sec:frameworks}
\subsection{HPX}
HPX is an asynchronous many-task runtime system (AMT).
It allows users to express the parallelism within their application by dynamically building a task-graph on-the-fly using HPX futures and continuations. For example, a function can be launched asynchronously using \texttt{hpx::async}, which returns an HPX future. This future can be used to suspend the current HPX task (\lstinline{get}) or define another task that will be automatically triggered once the original function is done (\lstinline{then}). Each task can subsequently spawn an arbitrary number of other tasks asynchronously and with little overhead, making parallel work available quickly, which is especially useful for parallelized tree traversals. Of course, there is a lot more functionality available. Particularly, HPX implements all C\texttt{++}23 APIs regarding parallelism and concurrency. For a full list of the functionality available, we refer to HPX's online documentation\footnote{\url{https://hpx-docs.stellar-group.org/latest/html/index.html}}. 
The tasks within HPX's task graph are then processed by a set of HPX worker threads. Thus, HPX can easily handle billions of tasks, which are processed by the worker threads as they become available within the graph.

HPX also contains various key features to enable the development of distributed codes~\cite{diehlparallel}.
For example, functions can also be launched asynchronously on other compute-nodes, still getting an HPX future back. There are also communication channels and other distributed features available that tie into the HPX task graph similarly. This allows users to build a distributed task graph.
This is \textcolor{black}{made possible} by HPX's Active Global Address Space (AGAS) and various networking backends (parcelports).

Overall, HPX is well suited to address the computational challenges of developing scalable, distributed AMR applications, directly addressing some of the challenges mentioned in Section~\nameref{sec:comp_problem}.
With \textcolor{black}{HPX}, users can exploit the parallelism within tree-structures more easily by expressing it in a similar graph structure, the HPX task-graph.
Through this task graph, which may contain both computation and communication tasks, users can automatically achieve an overlapping of computation and communication. \textcolor{black}{This helps one to} scale their codes to numerous compute nodes with a high parallel efficiency. Furthermore, thanks to some of our integrations, we can also incorporate accelerator devices into this task-graph. \textcolor{black}{This helps to} keep a high utilization of the accelerator devices by better overlapping their computations with the CPU tasks and communication.

\textcolor{black}{Use of} HPX can also help with portability.
HPX's portability is achieved through an interface (API) fully aligned with the latest C\texttt{++} standard specification. The API decouples the application from the upper layers of the runtime system and hides the low-level details of the underlying architecture. HPX provides general-purpose building blocks (such as C\texttt{++} standards conforming parallel algorithms), as well as high-level utilities coordinating asynchronous execution, and offers a programming abstraction that facilitates the programming of both shared-memory and distributed-memory systems through a uniform programming interface, scaling up from hand-held devices to HPC clusters. 

In terms of networking portability, HPX's modular architecture abstracts the concrete networking hardware allowing for Octo-Tiger to be completely independent of it. HPX supports TCP/IP, MPI, LCI~\cite{jiakun2023hpxlci,lci1_7}, OpenSHMEM~\cite{Poole2011}, and GASNet~\cite{bonachea2018gasnet} as possible networking conduits. In this work, we focus on two backends (parcelports): MPI and LCI. 

While HPX does not support developing portable GPU compute kernels, it directly supports integrating existing kernels into the generated asynchronous execution flow by launching those kernels as tasks that are part of the generated HPX task-graph. 

\subsection{Kokkos}
Kokkos implements a programming model in C\texttt{++} for writing performance portable applications targeting all major HPC platforms. For that purpose, it provides abstractions for both parallel execution of code and data management. Kokkos is designed to target complex node architectures with N-level memory hierarchies and multiple types of execution resources. It currently can use CUDA, HIP, SYCL, HPX, OpenMP, and C\texttt{++} threads as backend programming models~\cite{kokkos_docs}.
With Kokkos, users can write their compute kernel once then run it on the correct device by using the Kokkos execution space for the device and the associated Kokkos memory space for data.

Furthermore, Kokkos also contains GPU-compatible SIMD types. On CPU, these can be instantiated to use the appropriate SIMD instructions and registers of the current CPU. On GPU, these can be instantiated to mere scalar values. This allows user to make \textcolor{black}{explicit} use of the SIMD \textcolor{black}{resources} offered by the CPU, yet retain \textcolor{black}{compatibility} with GPUs.

These features help to decouple user code from specific hardware, greatly improving the portability.

\subsection{Octo-Tiger in a Nutshell}

\subsubsection{Octo-Tiger's Domain Science Use-Case: }
Octo-Tiger is a distributed, full 3D, multi-scale, multi-model, adaptive mesh-refinement (AMR) astrophysics code based on FVM methods \textcolor{black}{and} designed to study stellar mergers as described in Section~\nameref{sec:astro_problem}. It has been highly optimized for execution on distributed computing systems and can take advantage of the state of the art in GPUs. This allows Octo-Tiger to produce models with the high levels of resolution required to conduct convergence studies and understand the processes related to the increased ${}^{16}\textrm{O}$ concentration in DWD mergers that lead to the formation of R Coronae Borealis stars. Octo-Tiger is a real-world application that has proven to solve pressing science domain problems in the field of simulating the merger of two white dwarf stars bound in a binary star system~\cite{diehl2021octo,diehl2024fugaku,10.1093/mnras/stab937}. 

The Octo-Tiger code has several features that make it particularly \textcolor{black}{well-}suited for modeling interacting binaries. The entire grid structure rotates with the initial orbital frequency of the binary. This reduces the velocity of the material relative to the grid, and therefore reduces the non-physical effects due to numerical viscosity. The code carefully conserves global quantities like linear momentum, angular momentum, and energy, which is important for modeling the initial phase of mass transfer as small violations in conservation of those quantities can significantly effect the simulation results. Octo-Tiger conserves these quantities in the hydrodynamics solver by matching fluxes across jumps in refinement levels. As is the case with most Fast Multipole Method (FMM) solvers, the FMM conserves linear momentum. We also developed a special technique that enables the FMM to simultaneously conserve angular momentum. While the hydrodynamics solver does not conserve angular momentum, conservation in the gravity solver enables Octo-Tiger to conserve energy in the rotating frame; otherwise this would only be possible in a non-rotating frame. The AMR grid allows for the extension of the spatial domain boundaries to orders of magnitude larger than the orbital separation. \textcolor{black}{This allows} Octo-Tiger to model the extended low density outflows that occur during the merger process. The AMR grid also enables higher resolution in the regions of interest, making convergence studies more feasible.

\subsubsection{Octo-Tiger's Components and Data-Structure: }
Octo-Tiger models binary star systems as self-gravitating fluids.
Hence, it uses two interleaved solvers:
It employs a Fast Multipole Method (FMM) gravity solver and a finite volume hydrodynamics solver.
A third, experimental solver for radiation is currently in active development. 
As Octo-Tiger uses a third-order Runge-Kutta integration scheme, each time step involves three iterations of the hydrodynamics solver.
As the FMM is modified to conserve angular momentum, each time step further involves actually six (instead of three) iterations of the FMM solver.

The solvers operate on an adaptive octree with the AMR focusing on the atmosphere between the stars\textcolor{black}{, which is} where the mass \textcolor{black}{is being} exchanged between \textcolor{black}{stars}.
Each tree node contains an entire sub-grid with ${8}\times{8}\times{8}$ ($512$) cells to improve the computational efficiency.
\textcolor{black}{Each} compute kernel \textcolor{black}{typically} operates on one such sub-grid (and its ghost layers) at a time. \textcolor{black}{As a result,} we deal with a large number of potentially concurrent compute kernels when traversing the tree. However, each of these kernels only processes a comparatively small compute load. In theory, we can further increase the size of the sub-grids at compile-time to compensate, thus increasing the compute load per individual kernel invocation for better efficiency. Yet, this would in turn negatively impact the adaptivity (as we would get a less refined grid given the same amount of overall cells), the scalability (as the sub-grids are the components we distribute onto the compute-nodes) and lastly, it would impact the runtime of the FMM (as it uses the tree-structure to avoid computations by approximations).

Hence, both the efficient distributed tree traversals and the efficient handling of the small compute kernels are key to Octo-Tiger's overall performance.
Given the diverse set of available hardware in currently relevant supercomputers (ranging from NVIDIA, AMD and Intel GPU to x86 and ARM CPUs), portability is also a concern.
As such,  we face similar computational problems within Octo-Tiger as those that are outlined in Section~\nameref{sec:comp_problem}.


\section{Resolving the Challenges of Scalable AMR Applications with HPX and Kokkos}
\label{sec:challenges}
Using HPX in \textcolor{black}{conjunction} with Kokkos presents the opportunity to easily develop portable, yet highly scalable, distributed codes, even when using irregular tree-based structures.
\textcolor{black}{For this reason}, we turned to both frameworks when developing Octo-Tiger and porting it to GPUs, especially given our small team of core developers with only one developer being available for the refactoring and porting effort.

However, we found out that there were some challenges and missing pieces we had to resolve first to make these frameworks work together efficiently and portably (across CPUs and GPUs) within an AMR application such as Octo-Tiger:
\begin{itemize}
\item We needed to improve the HPX-Kokkos interoperability, avoiding both the blocking of HPX worker threads with Kokkos fences and avoiding conflicting thread pools for host execution.
\item We needed to address the issue of GPU device starvation caused by compute kernels that are too small (in turn, caused by the small workloads per tree-node in AMR applications such as Octo-Tiger). We furthermore needed to resolve the memory allocation overheads caused by temporary (but necessary) GPU buffers.
\item We needed additional SIMD types to portably and efficiently target A64Fx and RISC-V CPUs.
\item While the MPI parcelport (networking backend) in HPX works well on a range of machines, it is worthwhile to have more alternatives available for the networking that can be interchanged by the user at runtime.
\end{itemize}

In this section, we outline how we addressed each of those concerns in our previous work.
Thus, we introduce the HPX-Kokkos integrations (intended to address the HPX-Kokkos interoperability)~\cite{9460406}, specialized allocators and executors (intended to address work starvation and memory overheads for temporary buffers)~\cite{10024622}, additional SIMD types (to enable portable vectorization for A64FX and RISC-V CPUs while maintaining GPU compatibility of the kernels)~\cite{10.1109/P3HPC56579.2022.00014} and the LCI parcelport (offering another networking choice within HPX)~\cite{jiakun2023hpxlci}.

Notably, all of the integrations, backends, and tools mentioned in this Section are not specific to Octo-Tiger and can be used in other HPX applications as well. Yet, in a final part of this section, we outline how these solutions are applied within Octo-Tiger during our efforts to port this application to Kokkos (and GPUs in general).

\subsection{Combining Kokkos and HPX}
We first integrated Kokkos and HPX in~\cite{9460406}. Notably, this required an integration both ways: HPX needs to be integrated with Kokkos to enable asynchronous Kokkos calls (kernel launches, deep copies) as part of its own task graph; Kokkos needs to be integrated with HPX to make use of HPX's thread pool for CPU execution.

The first integration  \textcolor{black}{resulted} in the HPX-Kokkos \textcolor{black}{interoperability library, which} enables the execution of asynchronous Kokkos Kernels (or deep copies) into the HPX task graph.
This makes it easy to define continuation tasks that should be triggered once a kernel is done executing, such as doing post-processing on the results or communicating them, easing the development of distributed code.
This also eliminates \textcolor{black}{the} need for a CPU thread to actively wait on some GPU kernel as long as there is some other work remaining in the queue. \textcolor{black}{This allows} the runtime to easily manage thousands of concurrent kernel launches, using hundreds of GPU streams and just a few CPU worker threads on each process (usually one thread per utilized core).

HPX-Kokkos itself is a thin compatibility layer: depending on which Kokkos execution space is used, it uses the appropriate HPX API integrations to get an HPX \textit{Future} for this execution space that represents the event of the encapsulated work being done executing. 
For instance, when using a Kokkos CUDA execution space, HPX-Kokkos will use the \texttt{get\_future} functionality of the underlying HPX-CUDA integration that is directly implemented within HPX using event polling. 
\textcolor{black}{Similarly}, it will use the appropriate integrations for the Kokkos HIP and Kokkos SYCL execution spaces as well (as HPX includes HIP and SYCL integrations which we developed previously~\cite{9460406, 10.1145/3585341.3585354}). HPX-Kokkos merely maps the calls correctly and provides convenience functions such as \texttt{deep\_copy\_async} or \texttt{parallel\_for\_async} that call their Kokkos equivalents (\texttt{deep\_copy} or \texttt{parallel\_for}) and immediately obtain a HPX \textit{Future} for those.

HPX-Kokkos primarily helps us to integrate the asynchronous Kokkos GPU kernels with HPX, providing benefits to the programmer (it is easy to handle concurrent GPU kernels and define what should happen with their results asynchronously) and improving the runtime (by eliminating active waiting on GPU results in synchronizing barriers or Kokkos fences).

The second integration implements the HPX execution space as part of Kokkos.
This Kokkos HPX execution space (included within the Kokkos framework itself) helps with the CPU execution by enabling Kokkos kernels to run directly on the HPX worker threads.
This again gives us two advantages: First, it eliminates the need for conflicting thread pools (as we would have when using the OpenMP execution space). Second, it allows us to finely tune how many HPX tasks each Kokkos kernel should be split into (and thus, how many CPU threads are maximally used for their execution). This can be beneficial when dealing with many concurrent kernels, as we do in Octo-Tiger. For instance, we can use more CPU threads for kernels that work on tree nodes higher up in the tree (where there is less parallel work available) to avoid starving the CPU threads during the tree traversals.

\begin{figure*}[tb]
    \centering
    \includegraphics[width=0.95\linewidth]{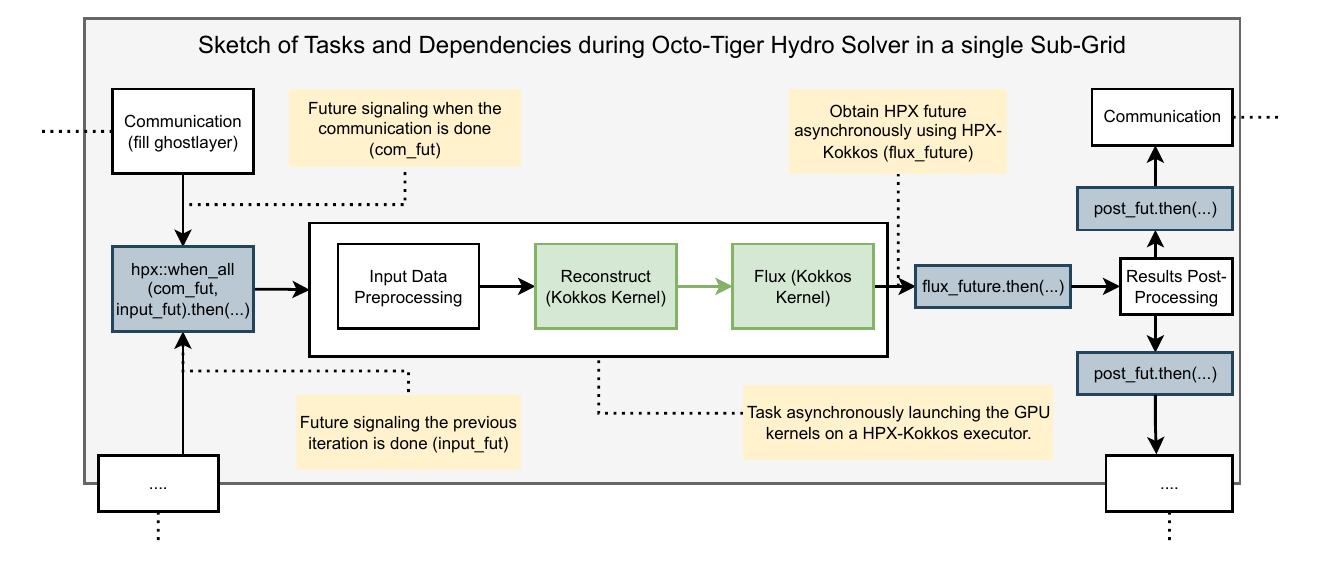}
    \caption{Illustration of a small part of Octo-Tiger's task-graph, depicting the tasks that are appended to the task-graph for every sub-grid in every iteration of Octo-Tiger's hydro solver.
Tasks (in white) are represented with HPX futures (for instance,  named \texttt{com\_fut}). These futures can be dynamically and asynchronously chained together into a task-graph using HPX functionality (in gray), such as \texttt{then} and \texttt{when\_all}. This allows us to build the task-graph on-the-fly whilst traversing Octo-Tiger's tree. With the HPX-Kokkos integration, it is possible to integrate asynchronous Kokkos kernels (in green) into this task-graph as well, freeing us from having to use blocking fences to process their results. Overall, this results in a multitude of concurrent CPU tasks, GPU tasks, and communication tasks. This leads to an interleaving of tasks, helping us to better utilize the available hardware resources and achieve scalability.}
    \label{fig:task_graph}
\end{figure*}

Using Kokkos and HPX together with these integrations allows us to both take full advantage of HPX (allowing us to finely interleave computation and communication tasks while transparently managing asynchronous execution graphs) and Kokkos (allowing us to target different CPU and GPU hardware with a single kernel implementation). Figure~\ref{fig:task_graph} shows an example of how a part of such a graph would look like for  Octo-Tiger.

\subsection{GPU-compatible SIMD Implementation with SVE Support}
While using Kokkos itself already allows us to target both CPU and GPU machines with the same kernel implementation, we still need to ensure that the kernel makes use of SIMD vectorization on the CPU.
Although relying on auto-vectorization might work in some kernels, the ones in Octo-Tiger contain too many branches, preventing the compiler from \textcolor{black}{automatically} vectorizing the code.
We can work around this problem by using SIMD masking. However, we still want to do this in a portable way as the code still needs to execute correctly on the GPU.

For this reason, we use C++ SIMD datatypes, specifically we use a common subset of the Kokkos SIMD types and the \texttt{std::simd} types (allowing us to decide at compile time, which of those to apply). As our work predates C++26 (and the official introduction of \texttt{std::simd}) we used the \texttt{std::experimental::simd} implementation within GCC (version 11 and upwards) during our development and testing. However, this is compatible with \texttt{std::simd}.
Both the Kokkos SIMD types and the \texttt{std::experimental::simd} overload all relevant operators and math operations, allowing us to use them just like normal floating point types. During compile time, they get either instantiated based on the relevant SIMD intrinsics (such as AVX512) or as scalar floating point types for kernels that run on the GPU. The types also support SIMD masks for conditional code branching.

We added the SIMD types and masks in all major Octo-Tiger compute kernels and tested the SIMD speedup in~\cite{10.1109/P3HPC56579.2022.00014} both with the Kokkos SIMD types and with \texttt{std::experimental::simd} types from GCC. Furthermore, as there were no \texttt{std::simd}-compatible types available for SVE, we implemented these types ourselves and tested them within Octo-Tiger. This allows us to also target the SIMD units in ARM processors (particularly, the A64FX CPUs used in the Supercomputer Fugaku). Eventually, our SVE SIMD type implementation from ~\cite{10.1109/P3HPC56579.2022.00014} has been adapted for inclusion into GCC to benefit the wider HPC community~\footnote{Merged into GCC in \url{https://github.com/gcc-mirror/gcc/commit/9ac3119fec81fb64d11dee8f853145f937389366}}.
Similar to the implementation of the SVE types, we also implemented SIMD types for RISC-V and its RVV SIMD instructions~\cite{diehl2024preparinghpcriscvexamining}. Unlike our SVE types, these have not been yet merged into GCC yet.

Together with the Kokkos HPX execution space, this allows us to have efficient CPU kernel implementations. Usually, we use rather fine-grained kernels where we utilize just 1 to 16 worker threads per kernel instead of the entire CPU. With many of those kernels running concurrently, it allows us to finely interleave computation and communication while at the same time utilizing the SIMD capabilities of the CPU to the best extent possible.

\subsection{Addressing GPU Starvation and Memory Overheads}
While fine-grained compute kernels (as the ones within Octo-Tiger) can be beneficial for adaptivity (more tree levels) and scalability (finer tree structure to distribute onto the compute nodes), it presents major problems for a GPU implementation:
On a GPU, we not only have more compute elements than on an average CPU, but \textcolor{black}{we also} rely on having enough work items to hide latencies and stalls. At worst, a too small GPU compute kernel with few work items will cause outright device starvation (where we do not even scale to all SMs on an NVIDIA A100). In addition, even when there is enough work for all streaming multiprocessors/compute units, the kernel will still run inefficiently if there are not enough concurrent work items available for latency hiding.

Typically, there are three ways to address this: increase the workload per kernel; increase the number of concurrent GPU kernels; or fuse kernels and run them as a single, larger GPU kernel. These strategies have their separate trade-offs: for instance, it is possible for Octo-Tiger to increase the workload per kernel by increasing the size of the sub-grid in each tree node. However, this negatively impacts adaptivity (if we aim for a similar grid size) and worsens the FMM performance (as we use the tree-structure to approximate the influence of far-away cells). Hence, it is best to combine it with the other two strategies: running multiple GPU kernels concurrently and employing dynamic GPU kernel fusion.

Handling concurrent GPU kernel launches is straightforward thanks to the integration of Kokkos kernel execution into the HPX task graph via HPX-Kokkos. \textcolor{black}{With this integration}, we can easily have $128$ concurrent kernels per process, even if the said process only contains a few worker threads to handle launches and communications. However, the overhead of creating a GPU stream (and thus a GPU executor) is significant. \textcolor{black}{To avoid this overhead,} we use a pre-allocated pool of GPU executors in each process, combined with a scheduler drawing from it on demand.

To enable the dynamic kernel fusion within HPX, we introduced a special executor in~\cite{10024622}. These executors (enabled by HPX \textit{Futures}) allow users to define code regions where multiple HPX tasks can cooperate by concurrently fusing their individual GPU kernels into one larger kernel (provided it is the same compute kernel operating on different data items). This reduces the overall number of GPU related API calls (fewer kernels and data transfers need to be initiated) and enlarges the work size of the fused kernel (better utilizing the GPU). This is especially valuable on GPUs that are less capable of efficiently running concurrent individual kernels.

We compared these three techniques to avoid GPU starvation and their impact within the Octo-Tiger hydro solver in greater detail in~\cite{10024622}. Here, we also compared the performance of the native CUDA/HIP kernels and their Kokkos counterparts on both NVIDIA A100 and AMD MI100 GPUs. 

Lastly, when dealing with many fine-grained compute kernels and associated communication ghost layers, the amount of required temporary GPU buffers can become a problem. Allocating and de-allocating them on-the-fly is too expensive but pre-allocating them for each tree node would significantly increase the memory requirements of Octo-Tiger. We avoid this problem by developing a dynamically growing memory pool for GPU device memory and for pinned memory on the host side. If we need a temporary GPU buffer, we can allocate it with a special allocator that will recycle a (currently unused) GPU buffer from the memory pool. Upon de-allocation, the buffer will simply go back into the pool of unused memory (but stays allocated from the perspective of the system). If there is no unused buffer of sufficient size available upon allocation, a new one will be created. This way, we only allocate as much memory for the temporary and communication buffers as is needed for maximum concurrency, allowing us to run larger scenarios with fewer compute nodes.

Both the kernel fusion functionality and the memory pool functionality are exposed through specialized executors and allocators which we published in the utility library CPPuddle.

\subsection{Inter-process Communication Optimizations}

HPX provides users with an Active Global Address Space in which processes can easily register and invoke remote functions and class methods (HPX \emph{actions}) on remote processes or (remote) global objects. This greatly simplifies the programming of Octo-Tiger as we do not need to deal with low-level message transfer. Meanwhile, building Octo-Tiger on top of HPX directly enables us to utilize various advanced HPX network layer techniques to speed up the inter-process communication of Octo-Tiger. This work focuses on two major high-performance network backends (\emph{parcelports}) in HPX: based on MPI and LCI.

The MPI parcelport deploys various techniques to enhance communication performance: arguments and return values of remote actions invoked are serialized into short control messages and optional large data messages to minimize memory copies; small messages are aggregated opportunistically to reduce message number; all the messages are sent/received with non-blocking MPI communication primitives to maximize communication overlaps.

Adding on top of the techniques used in the MPI parcelport, the LCI parcelport~\cite{jiakun2023hpxlci} further employs many communication features that do not exist in current standard MPI, including \textit{(a)} better communication primitives for active message style communication to eliminate tag matching, ordering, and memory copy overhead; \textit{(b)} lock-free completion queues for efficient polling of a large number of pending communication operations; \textit{(c)} lightweight interaction with and replication of low-level communication resources to expose more parallelism from the \textcolor{black}{hardware} level. Together, these techniques enhance the HPX communication layer with the ability to more efficiently handle a larger number of (possibly small) concurrent communications in a multi-threaded environment. 

\subsection{Application of our Solutions within Octo-Tiger}

Our Octo-Tiger application code was written such that it benefits from all aforementioned innovations. It was built from the ground using HPX, thus directly benefits from asynchronous task dependency construction and scheduling as well as intrinsic overlapping of computation and communication embedded in HPX's programming model. Each grid cell of its 3D computational mesh forms a tree-node in the spanned octtree and is represented by one HPX component that can be placed on any compute node. Using HPX's unified syntax for local and remote function calls for these components, we implemented efficient, distributed tree-traversals with little effort.

While Octo-Tiger was originally developed for CPU\textcolor{black}{-based} supercomputers, we gradually ported its computational hotspots to GPUs. This required some major refactoring of the code (requiring changes to some data-structures, like moving to Struct-Of-Arrays for some of them, or moving from an interaction list to a stencil approach for others). Given our small core developer team, this effort was done by a single developer and eventually turned Octo-Tiger into a GPU-accelerated application.
However, it is not yet a GPU-resident \textcolor{black}{code}: Only the computational-intensive kernels were ported to Kokkos and thus to GPUs, the rest of the application still runs on the CPU. While this already helps us to achieve major GPU speedups with Octo-Tiger~\cite{daiss2019piz, 9460406, diehl2021octo, 10024622}, it also means that the data is still primarily located on the main memory, not the GPU memory. Thus, there are numerous parts with CPU-GPU communication as the execution flow goes from GPU back to CPU execution, which puts more stress on the CPU-GPU connection and the main memory bandwidth itself, limiting the overall performance achieved by Octo-Tiger. Although there is more work to be done on Octo-Tiger, the existing GPU kernels already speed up the application noticeably in terms of total application runtime. In the context of this work, the frequent CPU-GPU data-transfers and the interleaving of numerous fine-grained CPU and GPU tasks actually provide a good stress test for our integrations and the HPX task-graph processing itself.
Regarding Octo-Tiger's GPU kernels themselves, the kernels are written using Kokkos and our HPX-Kokkos integration, ensuring portability. For high efficiency on CPU platforms, we utilize C\texttt{++} SIMD types within those Kokkos kernels. For instance, we use our SVE types on A64FX systems or scalar types when compiling for GPU platforms.
Thanks to the HPX-Kokkos integration, we are able to integrate the asynchronous Kokkos kernels into the HPX task-graph. This enables us to easily handle dozens of concurrent kernel launches per HPX worker thread, as we do not need to keep track of their status ourselves and let the HPX runtime handle this instead, also triggering the subsequent tasks once a Kokkos kernel is done. This automatically gains us the ability to interleave GPU kernels, CPU tasks, CPU-GPU memory transfers, and inter-node communication using the HPX task-graph. In our experience, the resulting automatic interleaving and overlapping of all of these tasks is key for achieving scalability at runtime. Notably, for a GPU-accelerated application, this is done without ever calling \texttt{Kokkos::fence} (or any equivalent \texttt{synchronize} method within CUDA); instead, we are relying on the event polling done by the HPX scheduling system.
On GPU platforms, we implement dynamic adaptive kernel fusion (aggregation) techniques to merge individual, concurrent kernel launches into a single GPU kernel to address the GPU starvation issues. For better memory utilization, we use the recycling allocators as described above.

We make use of the various networking backends offered by HPX that have been improved for this work to reduce networking overheads. We don't have to modify or even recompile the Oct-Tiger code in order to switch networking backends, which makes the performance tuning and comparison seamless.

While all of our contributions were developed in the context of Octo-Tiger, those are completely general and usable in other applications. However, Octo-Tiger serves as a perfect example of the usability of the whole software ecosystem and our contributions to it.

Despite the small Octo-Tiger core developer team, we were able to leverage these innovations to create a portable, highly adaptive simulation software, able to scale to thousands of GPU nodes whilst reducing the runtime per time step below 100 ms even for larger scenarios.

\section{How Performance Was Measured}
\label{sec:measure}

\subsection{Tools and Utilized Scenario}

\begin{table}[tb]
    \centering
    \rowcolors{2}{gray!25}{white}
    \caption{Average floating point operations (FLOP) per timestep measured over ten timesteps using the tool \textit{perf} on an Intel Skylake CPU, Number of cells, the memory usage, and the file size of the input file for all three refinement levels.}
    \label{tab:data:set}
    \begin{tabular}{c|llll}\toprule
     Level & FLOP &  \# cells & Memory & File size   \\\midrule
     10 & \num{1.68309E+12}  &   \num{3.8} M &   \num{11} GB &    \num{548} M \\
     11 & \num{1.74806E+13}  &  \num{40.2} M &  \num{113} GB &  \num{5.8} GB \\
     12 & \num{1.07915E+14}  & \num{257.3} M &  \num{724} GB & \num{32} GB \\\bottomrule
    \end{tabular}
\end{table}

Table~\ref{tab:data:set} details the Octo-Tiger scenarios we use, including the number of cells, the memory usage, and the input file size for all three levels. We use a state obtained from the production runs~\cite{shiber2024mnras} close to the merger for all three levels (10, 11, and 12) as a restart file. This allows us to use a very unbalanced mesh with adaptive mesh refinement around the two stars and where the aggregation belt will start, see Figure~\ref{fig:octo-tiger-3d-mesh}. We used timers in Octo-Tiger to obtain run-time values. However, our timers do not include the I/O time, hence all reported numbers are without I/O.

To make our performance results more comparable with other codes, we include an approximation of the FLOP/s that we obtain during the runs.
For this approximation, we obtain the total number of FLOPs for the scenario as follows:
Table~\ref{tab:data:set} shows the average floating point operations per time step over ten timesteps measured with the tool \textit{perf} on an Intel Skylake CPU. We compiled Octo-Tiger (using Spack) with no vectorization support and used double precision. These obtained FLOP/s are the basis for all runs on all platforms. The meshes of level 10 and the initial mesh of level 11 fit in the memory of a single node. All other meshes required distributed runs and we measured the FLOP on a single node and multiplied it with the number of nodes. 
This does not accommodate the unbalanced distribution of work. However, as we distribute the workload on as few compute nodes as possible to accommodate for the memory demands, the individual workload per compute node is large enough to make this a good enough approximation for our purposes nonetheless. This approximation is not exact, but provides us with a first estimate of our achieved FLOP/s.

We combined the measured FLOP with perf and the timers to calculate the FLOP/s. In the HPC community are two benchmarks to rank supercomputers HPCG and HPL, respectively. HPL ranks the supercomputer's efficiency in solving dense linear equation systems. HPCG uses sparse data structures that have low compute-to-date movement ratios concerning HPL. To compare the obtained FLOP/s, we use the HPCG~\cite{heroux2013hpcg} benchmark results from November 2023. We chose the HPCG benchmark due to its nature of streaming data and executing compute kernels, which aligns more closely with our application. Octo-Tiger adaptively refines the grid and traverses the oct-tree, which includes additional synchronization overhead and increased irregularity not present in the HPCG benchmark.

\begin{table}[tb]
    \centering
    \caption{HPCG results~\cite{heroux2013hpcg} from November 2023}
    \label{tab:overview:system}
    \rowcolors{2}{gray!25}{white}
    \begin{tabular}{lcc} \toprule
    System  & \makecell{HPCG \\ (PFLOP/s)} & \makecell{Frac of Peak \\ \%} \\\midrule
     Supercomputer\ Fugaku  & \num{16} & \num{3.0}\\
    Perlmutter   & \num{1.9} &  \num{2.4} \\
    Frontier    & \num{14} & \num{1.2}\\\bottomrule
    \end{tabular}
\end{table}

\subsection{Systems and Environments}

The Perlmutter supercomputer hosted at Lawrence Berkeley National Laboratory is a heterogeneous system comprised of AMD EPYC CPUs and NVIDIA A100 GPUs. It has $1536$ 40GB GPU nodes, $256$ 80GB GPU nodes, and $3,000$ CPU nodes connected by a 3-level dragonfly topology. We used only the GPU nodes for this study. Each GPU node is comprised of a single AMD CPU with 64 cores and 256GB of RAM and four A100 GPUs with 40GB/80GB of HBM per GPU. Each GPU has 4 HPE Slingshot 11 Network Interface Cards (NICs) and each pair of GPUs is connected via twelve 3rd generation NVLINKs with a speed of 25GB/s per NVLINK.  

The exascale Frontier supercomputer hosted at Oak Ridge National Laboratory is a heterogeneous system featuring AMD EPYC CPUs and AMD MI250X GPUs. It has $9,408$ HPE Cray EX235a nodes connected by a dragonfly topology. Each node features one 64-core AMD EPYC 7A53 ``Optimized 3rd Gen EPYC'' CPU, four AMD MI250X GPUs, each with 2 Graphics Compute Dies (GCDs), $64$ GB of high-bandwidth memory (HBM2E) on each GCD, and $512$ GB of DDR4 memory. The CPU is connected to each GCD by AMD's Infinity Fabric, which delivers up to $36+36$ GB/s. Each GCD on a node is interconnected by AMD's Infinity Fabric, which delivers up to $50+50$ GB/s for GCDs across GPUs and up to $200+200$ GB/s for GCDs on
the same GPU. Each node is connected to the network by 4 Slingshot 11 NICs.

The supercomputer\ Fugaku is a massively parallel computer system with $158,976$ nodes based on A64FX~\cite{9355239} cores.
The A64FX is based on the Armv8.2 architecture with scalable vector extensions (SVE). Each processor has four groups of cores called Core Memory Group (CMG) connected to a ring-bus network. Each CMG contains 12 computation cores and 1 or 0 assistant cores. A computation core has 64KiB L1I and 64KiB L1D caches. The cores in a CMG share an 8MiB L2 cache and 8GiB HMB2 memory. The total size of the memory in a node is 32GiB and the aggregate memory bandwidth is 1024GB/s. The network of Fugaku is called Tofu Interconnect D (Tofu-D), which is a hybrid 6D mesh/torus network of 40.8 GB/s total injection bandwidth per node~\cite{8514929}. Ookami is a small NSF-funded research cluster at Stony Brook University with 174 nodes containing A64FX CPUs and $32$ GB RAM. All nodes are connected with InfiniBand HDR 200. Except for the network architecture, it is similar to the supercomputer Fugaku. Darwin is a research testbed cluster funded by the Computational Systems and Software Environments (CSSE) subprogram of LANL’s ASC program (NNSA/DOE). 
It is a very heterogeneous cluster with a wide variety of hardware available, including x86, Power PC, and ARM CPU architectures, systems with terabytes of memory, and a variety of GPUs and other accelerators. For this work, we used only the NVIDIA Grace-Hopper partition of the machine. The MILK-V Pioneer is the first workstation-grade development machine for exploring RISC-V. Containing a Sophon SG2042 64-core RISC-V CPU with a 2 GHz clock frequency, 128 GB DDR 4 memory, and a 1TB PCIe 3.0 SSD. Table~\ref{tab:compilers} summarizes the used compilers and environments. The data artifact is available on GitHub\footnote{\url{https://github.com/STEllAR-GROUP/OperationBell24}}.

\begin{table*}[tb]
    \centering
    \rowcolors{2}{gray!25}{white}
    \begin{tabular}{l|lllllll}\toprule
     System & compiler & mpi & hpx & kokkos & hpx-kokkos & GPU  SDK & lci \\ \midrule
     Fugaku  & gcc 12.2.0 &  fujitsu-mpi head  & 1.9.1 & 4.0.01 & 0.4.0 & -- & -- \\
     Perlmutter & gcc 12.3.0 & cray-mpich 8.1.28 & \textit{310f449} & 4.0.01 & 0.4.0 & CUDA 12.2 & \textit{6dcb6a8} \\
     Frontier & rocmcc 5.7.0 & cray-mpich 8.1.28 & \textit{310f449} & 4.0.01 & 0.4.0 & HIP 5.7.0 & --\\
     Ookami & gcc 12.2.0 & openmpi 4.1.5 & \textit{40eef48} & 4.1.00 & 0.4.0 & -- & \textit{6dcb6a8}\\\bottomrule
    \end{tabular}
    \caption{Software environments used on the systems. Some systems required patches and we provided the short git hash in italic.  }
    \label{tab:compilers}
\end{table*}

\section{Performance Results}
\label{sec:results}
In this section we show the scalability and performance of Octo-Tiger.
We will use four platforms: Perlmutter, Frontier, and Fugaku as the major ones; Ookami as an A64FX machine compatible with LCI.
First, we focus on the scalability difference between the two relevant HPX networking backends for inter-process communication (LCI and MPI parcelports). Here, we limit ourselves to Perlmutter and Ookami (as we could not test LCI on Fugaku due to its incompatibility with the Tofu-D interconnect nor on Frontier due to time limits).
Afterward, we follow up by showing the FLOP/s (based on our approximation of the flops needed by the scenario) and cell-processed per second (to stay comparable to our previous results).

\subsection{Alternative Communication Backend}

To analyze the effects of using different HPX parcelports at scale, we ran the largest scenario (level 12) on Perlmutter going from $32$ nodes up to a full system run.
For the full system run, we achieve a total time of $5.61$s using the LCI parcelport and $9.7$s when using the MPI parcelport, gaining us a speedup of $1.73$x with LCI. The parallel efficiency with respect to the smallest run with $32$ nodes is about $42$\% with the LCI and $35$\% with the MPI parcelport.
Overall, on Perlmutter the LCI parcelport consistently performs better than the (older) MPI parcelport within HPX. For the full system runs, we achieve a speedup of $1.73$x using the LCI parcelport over the MPI one.
From the perspective of Octo-Tiger core developers, this is especially beneficial since we can reap this LCI speedup without having to modify anything about Octo-Tiger (as the networking backend within HPX is abstracted away).

\begin{figure}[tb]
    
    \centering
    \includegraphics[width=0.95\linewidth]{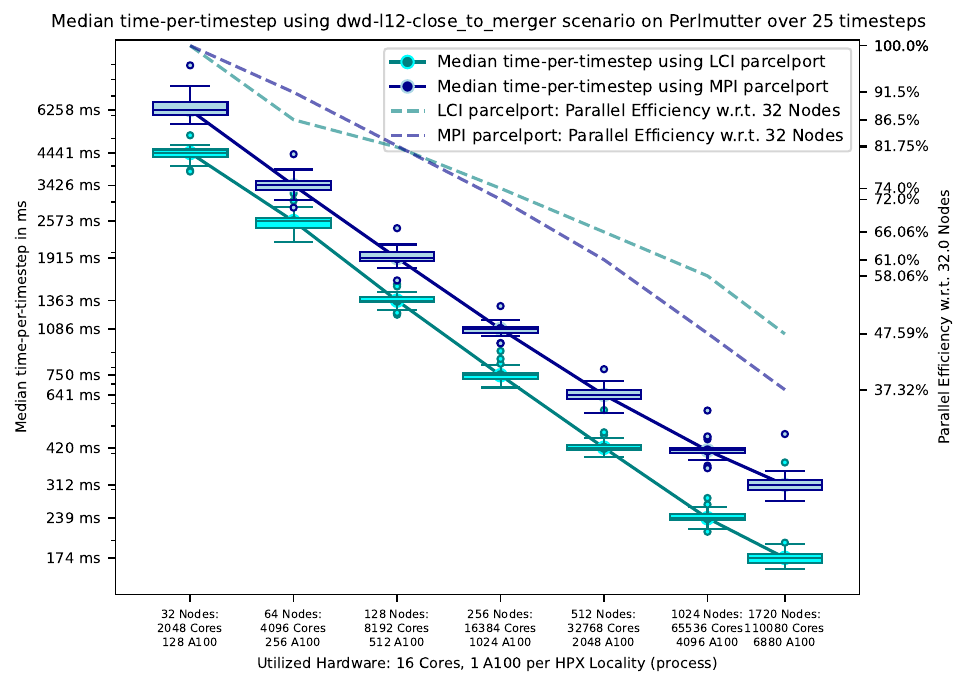}
    
    \caption{Median runtime per computational timestep and parallel strong scaling efficiency of the kernel execution per timestep on Perlmutter comparing the use of the MPI and LCI HPX parcelports}
    \label{fig:timestep_perlmutter}
\end{figure}

This Octo-Tiger total time was measured over 25 time steps on Perlmutter ($10$ time steps on Ookami). As production run scenarios usually take tens of thousands of time steps, we show the scaling in terms of the median runtime per timestep in Figure ~\ref{fig:timestep_perlmutter} (with a slightly better parallel efficiency of $47.6$\% than for the total time).
Here, it is  worth highlighting that this runtime per time step (including all computations and communication for the multiple, required solver iterations) just takes about $174$ms. Here, we benefit massively from our fine interleaving of computation and communication.

The outliers in Figure~\ref{fig:timestep_perlmutter} are worth mentioning: Here, we can see the effect of the dynamic GPU memory pools that Octo-Tiger uses. If a GPU buffer is required, the system will try to draw one from these pools and only create one if none is currently available. During the first time step, there are no buffers in the pool yet, hence it is filled over time during this first time step.

\begin{figure}[tb]
    \centering
    \includegraphics[width=0.95\linewidth]{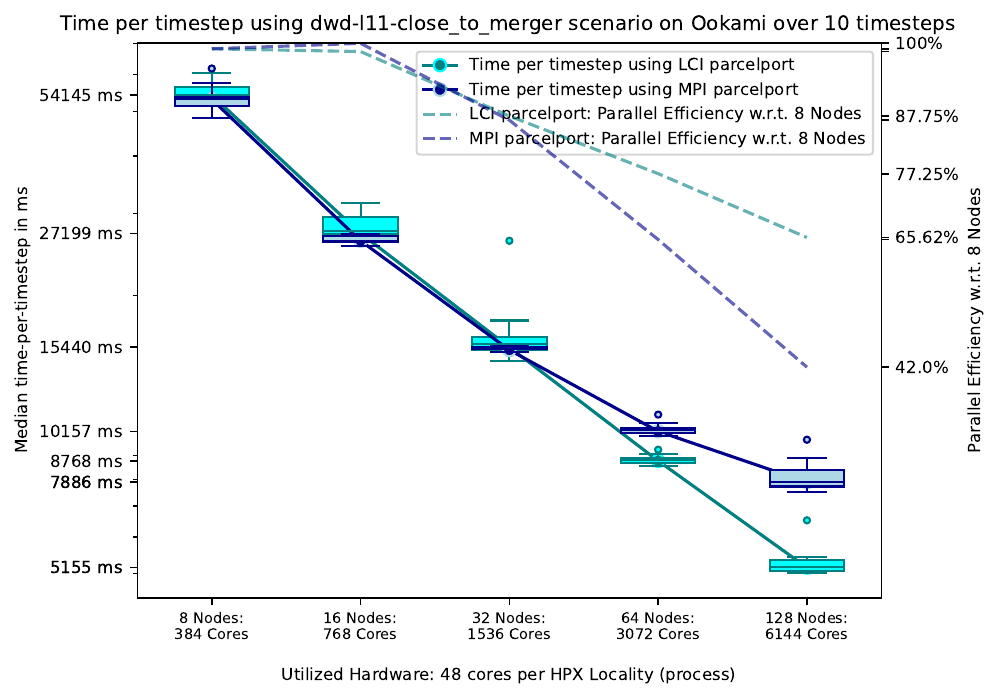}
    
    \caption{Total Runtime and parallel strong scaling efficiency on Ookami comparing the use of the MPI and LCI HPX parcelports.}
    \label{fig:total_time_ookami}
\end{figure}

We ran the smaller (level 11) scenario with $10$ timesteps on the A64FX Ookami machine as well. While the machine is a lot smaller than Perlmutter, it allowed us to test the runtime with both MPI and LCI parcelports on an A64FX platform as well (which we could not do on Fugaku as LCI is not yet compatible with the Tofu-D interconnect). The results can be seen in Figure~\ref{fig:total_time_ookami}.
While the MPI parcelport performs slightly better than its LCI counterpart at smaller node counts, the LCI parcelport achieves significantly better performance at higher scales (starting with $64$ nodes).
Here, we achieve a LCI speedup of $1.53$x when using LCI over MPI with $128$ nodes.

\subsection{Large-Scale Runs}

The previous section demonstrated the scalability, low runtime per time step, and the speedup we can achieve simply by switching to the LCI parcelport.
In this section, we show the total FLOP/s Octo-Tiger can achieve when performing large-scale runs on three major platforms. We use the estimated FLOP/s for each scenario in Table~\ref{tab:data:set}.
Overall, the simulation is not compute-bound as Octo-Tiger has to deal with adaptive mesh refinement, ghost layer exchanges for each sub-grid, and many smaller auxiliary functions that need to iterate over the data. Furthermore, data-structure conversions within Octo-Tiger allowing the CPU-only parts of the code to interact with the GPU-accelerated code are required to ensure best possible performance on both, the CPUs and GPUs.
In general, the workloads of Octo-Tiger are more irregular and fine-grained than those in the HPCG benchmark. It is expected that Octo-Tiger cannot achieve a significant fraction of the total peak performance of the underlying systems.

\begin{figure}[tb]
    \centering
    \includegraphics[width=0.95\linewidth]{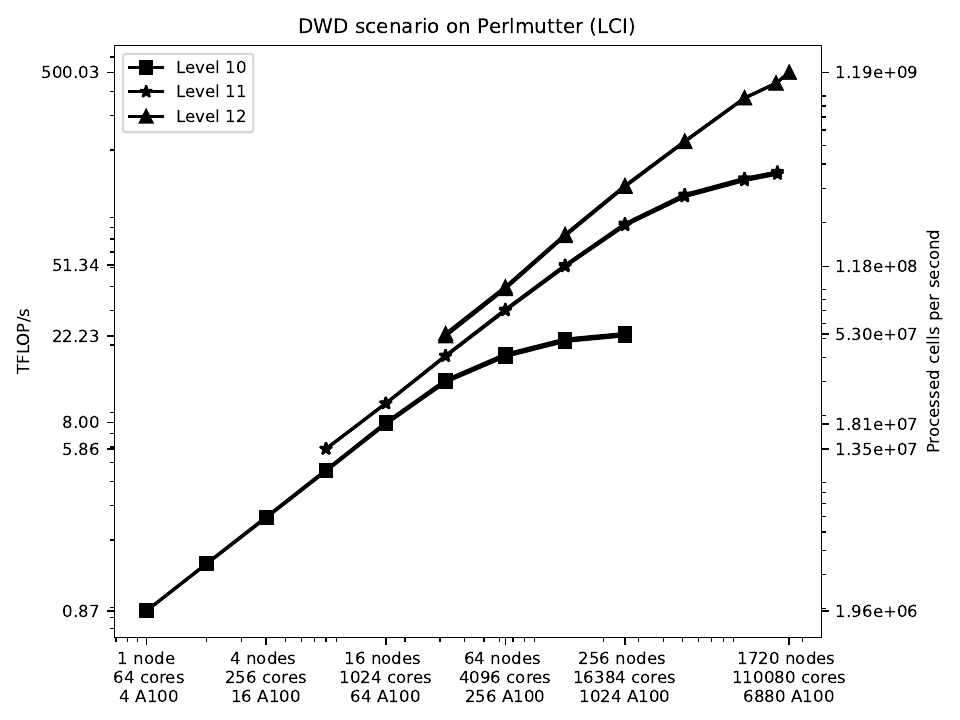}
    \caption{TFLOP/s (left axis) and number of processed cells per second (right axis) on Perlmutter using LCI for three levels of refinement (level 10, 11, and 12).}
    \label{fig:distributed:perlmutter}
\end{figure}

Figure~\ref{fig:distributed:perlmutter} shows the achieved TFLOP/s ($10^{12}$ FLOP/s) and the number of processed cells per second on Perlmutter. Note that we show the number of processed cells per second to make the runs on Perlmutter comparable with Frontier and supercomputer Fugaku. We show the LCI results here due to the better performance. On Perlmutter, we scale up to $256$ nodes with level 10 having around $3,711$ cells per GPU per time step. For level 11, we scale up to $1,475$ nodes with an average of $6,814$ cells per GPU per time step. \textcolor{black}{Though Perlmutter features $1,536$ 40 GB GPUs, only} $1,475$ were available during our reservation for the full system run. Combining the 40 GB GPUs and the 80GB GPUs with a higher memory bandwidth allowed for the largest run of $1,720$ nodes. Here, each GPU had on average $37,398$ cells per GPU per time step. 

\begin{figure}[tb]
    \centering
    \includegraphics[width=0.95\linewidth]{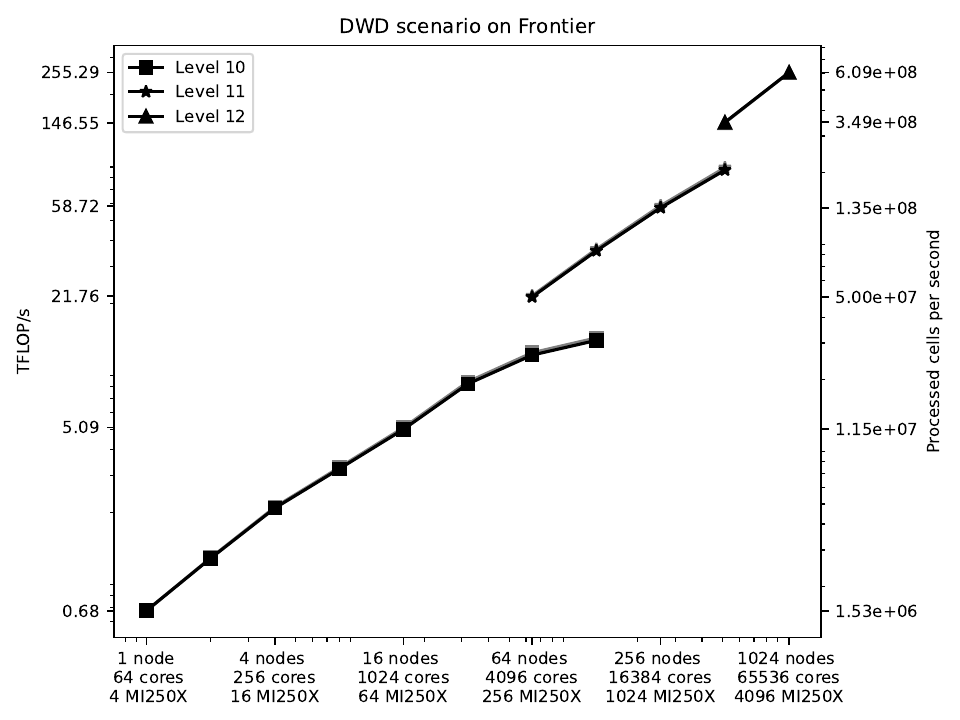}
    \caption{TFLOP/s (left axis) and number of processed cells per second (right axis) on Frontier for an increasing amount of utilized nodes for three levels of refinement (levels 10, 11, and 12).}
    \label{fig:distributed:frontier}
\end{figure}

Figure~\ref{fig:distributed:frontier} shows the achieved TFLOP/s and the number of processed cells per second on Frontier. Level 10 scales up to $128$ nodes with an average of $7,422$ cells per GPU per time step. Level 11 scaled up to $512$ nodes with 
an average of $19,629$ cells per GPU per time step. Level 12 runs finished up to $1,024$ nodes with an average of $62,817$ cells per GPU per time step.
On Frontier, for the level 11 scenario, we get a total time $20.89$s with $64$ nodes. Going to $512$ nodes, we get a total time of $5.08$s, gaining us a parallel efficiency of $51.38$\% (regarding the run smallest level 11 run with $64$ nodes).
While the scalability still looks good (especially for the larger level 12 scenario), we were not able to go beyond $1,024$ nodes due to the queue waiting times leading up to the submission date.

\begin{figure}[tb]
    \centering
    \includegraphics[width=0.95\linewidth]{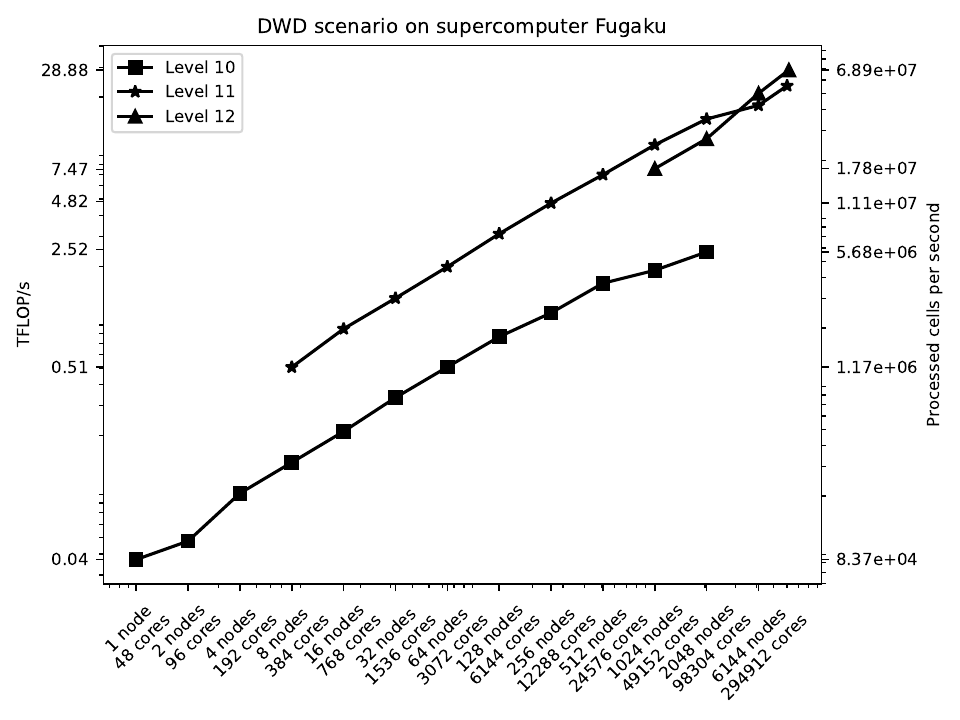}
    \caption{TFLOP/s  (left axis) and number of processed cells per second (right axis) on supercomputer\ Fugaku for three levels of refinement (levels 10, 11, and 12).}
    \label{fig:distributed:fugaku}
\end{figure}

Figure~\ref{fig:distributed:fugaku} shows the achieved TFLOP/s and the number of processed cells per second on supercomputer\ Fugaku for all three levels. We used the SVE backend for our SIMD types. Level 10 scaled from a single node up to 2048 nodes with around $40$ cells per core per time step. Level 11 fit in $8$ nodes and scaled up to $6,000$ nodes with around $140$ cells per core per time step. Level 12 runs from $1,024$ nodes up to $6,144$ nodes with around $872$ cells per core per time step. We need to mention here that going from level 11 to level 12 the number of cells and the memory usage increased by $6.4$, see Table~\ref{tab:data:set}. Note that the largest run was on $294,912$ cores. Here, we reached a total time $97.14$s with a parallel efficiency $64.5$\% (with respect to the smallest level 12 run using $1,024$ nodes). A notable difference is that the Fujitsu MPI used on supercomputer\ Fugaku only supports \lstinline[language=c++]{MPI_THREAD_SERIALIZED} and we had to add additional synchronization in our code. However, we could use \lstinline[language=c++]{MPI_THREAD_MULTIPLE} for the other systems. The additional synchronization overhead shows up for level 12 with many more messages. For a detailed study on the MPI threading level, we refer to~\cite{diehl2024fugaku}. We observe fewer TFLOP/s for level 12 than for level 11 on A64FX, which we do not observe on Frontier and Perlmutter. For this reason, we have not performed runs on even larger node counts on supercomputer\ Fugaku.

We are not sure why there is occasional, although rare, superlinear scaling in Figure~\ref{fig:distributed:fugaku} as this is not cache related (otherwise the superlinear scaling would occur regularly, which it does not). We suspect this was caused by the current network utilization on Fugaku during the time of our runs, but we cannot say for certain.

Table~\ref{tab:distributed:flops} shows the achieved TFLOP/s for the largest node count on each level of refinement. We compare the obtained TFLOP/s with the peak performance reported by the HPCG benchmark, see Table~\ref{tab:overview:system}, and report the percentage. 

\newcommand{\flopnumber}[1]{\num[round-mode=places,round-precision=2]{#1}}
\begin{table*}[tb]
    \centering
    \rowcolors{2}{gray!25}{white}
    \caption{TFLOP/s on the largest node count for three refinement levels and the corresponding percentage of HPCG peak performance in parentheses. Note that HPCG numbers are reported for full system runs which is shown here for level 12 on Perlmutter. For all other runs, we scaled the HPCG full run to the corresponding node count.}
    \label{tab:distributed:flops}
    \begin{tabular}{l|llll}\toprule
     Machine/Level&   10 & 11 & 12  \\\midrule
     Frontier & \flopnumber{13.1511955} ($6.93$) & \flopnumber{85.87595379} ($22.62$) & \flopnumber{255.2876523} ($16.81$) \\
     Fugaku & \flopnumber{2.516384853} ($1.22$) & \flopnumber{24.10278195} ($3.99$) & \flopnumber{28.88332774} ($4.67$) \\
     Perlmutter (MPI) & \flopnumber{12.35952168} ($9.08$) & \flopnumber{86.78939895} ($6.53$) & \flopnumber{278.0700959} ($15.20$) \\
     Perlmutter (LCI) & \flopnumber{20.61165458} ($15.14$) & \flopnumber{130.730089} ($9.83$) & \flopnumber{480.8000756} ($26.28$) \\\midrule
    \end{tabular}
\end{table*}

We observed the lowest TFLOP/s on supercomputer Fugaku compared to the two machines with GPU acceleration. 
The full system run on Perlmutter using mixed GPUs achieved almost twice the number of TFLOP/s ($500.03$ TFLOP/s) as the run using $1024$ Frontier nodes ($255.29$ TFLOP/s).

Table~\ref{tab:performance:risc} shows distributed scaling on up to two MILK RISC-V nodes for refinement level 10. We observe an improvement of a factor of around $1.7x$ going from one node to two nodes. Currently, only two MILK RISC-V nodes are available to us and we hope to do larger runs on future clusters.

\begin{table}[tb]
    \centering
    \caption{GFLOP/s and the number of processed cells per second on MILK RISC-V nodes.}
    \label{tab:performance:risc}
    \rowcolors{2}{gray!25}{white}
    \begin{tabular}{c|ll}\toprule
     Nodes &  GFLOP/s & Processed cells per second \\\midrule
     1    & \num{2.65} & \num{599061}\\
     2   &  \num{4.55} & \num{1028090} \\\bottomrule
    \end{tabular}
\end{table}

Additionally, we have performance results running on NVIDIA Grace Hopper on an early access node. However, only node-level results are available, and distributed runs on LANL's Venado using the finalized hardware and software stack are planned. Node-level scaling on Intel Xeon Max CPUs and Intel Max GPUs are available and runs on Aurora are targeted.


\section{Conclusion and Outlook}
\label{sec:conclusion}
The simulation of large time-dependent, three-dimensional, multi-scale, multi-model physical systems based on dynamic and adaptive meshes poses grand challenges to the fastest supercomputers. In this work, we have pioneered the simulation of stellar mergers on massively parallel systems with heterogeneous hardware and made significant contributions to both the application domain and high-performance computing methodology. 

We have created Octo-Tiger, an astrophysical code that can simulate astrophysical phenomena such as the merger of binary star systems. Exemplary for many applications, we need to employ dynamic adaptive mesh refinement to efficiently bridge across multiple physical scales. In contrast to alternative approaches based on smooth particle hydrodynamics, a mesh-based finite volume scheme \textcolor{black}{is used to} allow us to conserve important physical quantities such as momentum and energy up to machine precision. Additionally, it enables convergence studies that are critical to a deeper understanding of the underlying phenomena. We realize, for the first time, highly resolved simulations with up to 257 million dynamically adaptive cells, which was previously infeasible. With \textcolor{black}{this work}, we have prepared the \textcolor{black}{groundwork} for extreme-scale simulations to gain a deeper understanding of the evolution of our universe.

Dynamic and adaptive meshes covering multi-physics (such as hydrodynamics and gravity) are, however, a clear mismatch with modern massively distributed and heterogeneous high-performance computers. Dynamic load balancing and code portability are but two of the grand challenges. 

We solve load balancing by employing asynchronous many task parallelism via the graph-based scheduling of HPX, Kokkos kernels, and lightweight, non-blocking communication resources.
We have pioneered the integration of Kokkos, HPX, and SIMD-types to achieve code portability across a spectrum of heterogeneous supercomputers including Perlmutter, Frontier, and Fugaku, encompassing all three major GPU architectures as well as x86, ARM, and RISC-V CPUs.

To care for the communication requirements of adaptive algorithms, asynchronous programming models, and heterogeneous systems, which are usually heavily multi-threaded with finer-grained, point-to-point, concurrent communication, we have contributed a seamless switch of the underlying communication libraries at run time. Besides classical MPI, we have shown the benefits of employing an alternative communication library (LCI) as the communication backend to significantly improve communication efficiency. 

In our largest runs \textcolor{black}{with $257$ million dynamically adaptive cells}, we have achieved $51.37$\% parallel efficiency on Frontier on $32,768$ cores and $2,048$ MI250X GPUs with about $17$\% HPCG peak performance; $64.47$\% on $294,912$ cores on supercomputer\ Fugaku ($5$\% HPCG peak); and $47.59$\% for a full system run on Perlmutter on $4,110,080$ cores, and $6,880$ A100 GPUs using both the available 40GB A100 nodes with the 80GB A100 nodes simultaneously ($26$\% HPCG peak). 

Our contributions have also brought astrophysical simulations into a unique position to gain code and performance portability for novel and future systems. On a node level, we \textcolor{black}{have} already shown code portability to the new Intel GPUs, to Raspberry PIs, and to RISC-V CPUs. These efforts are pivotal in
laying the groundwork for its use on tomorrow's upcoming ARM64 and RISC-V HPC systems such as the European RISC-V flagship supercomputer announced for 2026.

\textcolor{black}{These} computational and algorithmic achievements are not specific to our guiding astrophysical application. The insights and significant improvements gained can be directly transferred to other mesh-based multi-physics codes with dynamic adaptivity, thus promising a high impact on future code developments.

However, while the scalability and portability results with Octo-Tiger look extremely promising, there is still much to improve within the application itself, as is shown by the achieved FLOP/s.
One major downside of the current version of Octo-Tiger, is that it is a GPU-accelerated application, not yet a GPU-resident one.
While its computational hotspots within the hydro and gravity solvers have been ported to Kokkos (and thus to GPUs), there are many parts of its code that still run on CPU. This means that the GPU results constantly need to be transferred to the CPU to be processed there by \textcolor{black}{other parts of code}.
Worse, as we had to refactor some utilized data-structures to make Octo-Tiger more suitable for GPUs (moving from Array-of-Structs to Struct-of-Arrays, for example, see:~\cite{pfander2018octotiger}), the interface between these older and newer parts sometimes includes data-structure conversions.
\textcolor{black}{This} puts a lot of stress on the main memory bandwidth, turning it into a bottleneck. This is mostly a result of refactoring and porting Octo-Tiger piece by piece.
Despite this bottleneck, we still achieve major speedups when using GPUs over CPUs \cite{9460406, 10.1145/3585341.3585354, 10024622, diehl2021octo}. However, the aforementioned legacy CPU code parts prevent us from realizing Octo-Tiger's full potential on GPU machines just yet.

Thus, our next steps regarding Octo-Tiger include finally removing these older parts of Octo-Tiger, focusing on the ones that include the data-structure conversion. Our goal is to eventually turn Octo-Tiger into a fully GPU-resident application, with all its main data-structures per sub-grid being located on the GPU.

\section*{Acknolegments}
This research used resources of the National Energy Research Scientific Computing Center (NERSC), a U.S. Department of Energy Office of Science User Facility located at Lawrence Berkeley National Laboratory, operated under Contract No. DE-AC02-05CH11231 using NERSC award DDR-ERCAP0028472. This research used resources of the Oak Ridge Leadership Computing Facility at the Oak Ridge National Laboratory, which is supported by the Office of Science of the U.S. Department of Energy under Contract No. DE-AC05-00OR22725. This research used computational resources of the supercomputer\ Fugaku provided by RIKEN Center for Computational Science. This work was supported by the U.S. Department of Energy through the Los Alamos National Laboratory (LANL). LANL is operated by Triad National Security, LLC, for the National Nuclear Security Administration of U.S. Department of Energy (Contract No. 89233218CNA000001). We also thank the LANL Advanced Simulation \& Computing Program and CCS-7 Darwin cluster for computational resources. The authors would like to thank Stony Brook Research Computing and Cyberinfrastructure, and the Institute for Advanced Computational Science at Stony Brook University for access to the innovative high-performance Ookami computing system, which was made possible by a \$5M National Science Foundation grant (\#1927880). The support we received from the Center of Computation and Technology at Louisiana State University was invaluable. The authors also acknowledge the technical support we received from NVIDIA (Scot Halverson) in the early stages of the project. Assigned: LA-UR-24-23457 (Rev. 3).

\bibliographystyle{SageV}
\bibliography{literature}

\end{document}